\documentstyle[preprint,aps]{revtex}

\begin{document}
\title{{\bf Exact solutions of the modified Kratzer potential plus ring-shaped
potential in the }$D${\bf -dimensional Schr\"{o}dinger equation by the
Nikiforov-Uvarov method }}
\author{Sameer M. Ikhdair\thanks{%
sikhdair@neu.edu.tr} and \ Ramazan Sever\thanks{%
sever@metu.edu.tr}}
\address{$^{\ast }$Department of Physics, \ Near East University, Nicosia, North
Cyprus, Mersin 10, Turkey\\
$^{\dagger }$Department of Physics, Middle East Technical University, 06531
Ankara, Turkey.}
\date{\today
}
\maketitle

\begin{abstract}
We present analytically the exact energy bound-states solutions of the
Schr\"{o}dinger equation in $D$-dimensions for a recently proposed modified
Kratzer potential plus ring-shaped potential by means of the conventional
Nikiforov-Uvarov method. We give a clear recipe of how to obtain an explicit
solution to the wave functions in terms of orthogonal polynomials. The
results obtained in this work are more general and true for any dimension
which can be reduced to the standard forms in three-dimensions given by
other works.


\end{abstract}


\section{Introduction}

\noindent The important task of quantum mechanics is to find the exact
bound-states solution of the Schr\"{o}dinger equation for certain potentials
of physical interest. Generally speaking, there are a few main traditional
methods to study the exact solutions of quantum systems like the Coulomb,
the harmonic oscillator [1,2], the pseudoharmonic [3,4] and the Kratzer
[4,5] potentials. Additionally, in order to obtain the bound-states
solutions of central potentials, one has to resort to numerical techniques
or approximation schemes. For many of the quantum mechanical systems, most
popular approximation methods such as shifted $1/N$ expansion [6],
perturbation theory [7], path integral solution [8], algebraic methods with
the SUSYquantum mechanics method and the idea of shape invariance, further
closely with the factorization mrthod [9], exact quantization rule [10,11],
the conventional Nikiforov and Uvarov (NU) method\ [12-25]. Some of these
methods have drawbacks in applications. Although some other methods give
simple relations for the eigenvalues, however, they lead to a very
complicated relations for the eigenfunctions.

The study of exact solutions of the Schr\"{o}dinger equation for a class of
non-central potentials with a vector potential and a non-central scalar
potential is of considerable interest in quantum chemistry [26-35]. In
recent years, numerous studies [36-40] have been made in analyzing the bound
states of an electron in a Coulomb field with simultaneous presence of
Aharanov-Bohm (AB) [41] field, and/or a magnetic Dirac monopole [42], and
Aharanov-Bohm plus oscillator (ABO) systems. In most of these studies, the
eigenvalues and eigenfunctions are obtained by means of seperation of
variables in spherical or other orthogonal curvilinear coordinate systems.
The path integral for particles moving in non-central potentials is
evaluated to derive the energy spectrum of this system analytically [43]. In
addition, the idea of SUSY and shape invariance is also used to obtain exact
solutions of such non-central but seperable potentials [44]. Very recently,
the conventional NU method has been used to give a clear recipe of how to
obtain an explicit exact bound-states solutions for the energy eigenvalues
and their corresponding wave functions in terms of orthogonal polynomials
for a class of non-central potentials [45].

Recently, Chen and Dong [46] found a new ring-shaped potential and obtained
the exact solution of the Schr\"{o}dinger equation for the Coulomb potential
plus this new ring-shaped potential which has possible applications to
ring-shaped organic molecules like cyclic polyenes and benzene. Very
recently, Cheng and Dai [47], proposed a new potential consisting from the
modified Kratzer's potential [48] plus the new proposed ring-shaped
potential in [46]. They have presented the energy eigenvalues for this
proposed exactly-solvable non-central potential in three dimensional $($%
i.e., $D=3)$-Schr\"{o}dinger equation through the NU method. The two quantum
systems solved by Refs [46,47] are closely relevant to each other as they
deal with a Coulombic field interaction except for a slight change in the
angular momentum barrier acts as a repulsive core which is for any arbitrary
angular momentum $\ell $ prevents collapse of the system in any dimensional
space due to the slight perturbation to the original angular momentum
barrier.

The conventional Nikiforov-Uvarov (${\rm NU}$) method [12], which received
much interest, has been introduced for solving Schr\"{o}dinger equation
[13-21], Klein-Gordon [22,23], Dirac [24] and Salpeter [25] equations. We
will follow parallel solution to [47] and give a complete exact bound-states
solutions and normalized wave functions of the $D$-dimensional
Schr\"{o}dinger equation with modified Kratzer plus ring-shaped potential, a
Coulombic-like potential with an additional centrifugal potential barrier,
for any arbitrary $\ell ^{\prime }$-states using the conventional
Nikiforov-Uvarov method. Our general solution reduces to the standard
three-dimensions given by Ref. [47] in the limiting case of $D=3$.

This work is organized as follows: in section \ref{BC}, we shall briefly
introduce the basic concepts of the NU method. Section \ref{ES} is mainly
devoted to the exact solution of the Schr\"{o}dinger equation in $D$%
-dimensions for this quantum system by means of the ${\rm NU}$ method.
Finally, the relevant results are discussed in section \ref{RAC}.

\section{Basic Concepts of the Method}

\label{BC}The NU method is based on reducing the second-order differential
equation to a generalized equation of hypergeometric type [12]. In this
sense, the Schr\"{o}dinger equation, after employing an appropriate
coordinate transformation $s=s(r),$ transforms to the following form:
\begin{equation}
\psi _{n}^{\prime \prime }(s)+\frac{\widetilde{\tau }(s)}{\sigma (s)}\psi
_{n}^{\prime }(s)+\frac{\widetilde{\sigma }(s)}{\sigma ^{2}(s)}\psi
_{n}(s)=0,
\end{equation}
where $\sigma (s)$ and $\widetilde{\sigma }(s)$ are polynomials, at most of
second-degree, and $\widetilde{\tau }(s)$ is a first-degree polynomial.
Using a wave function, $\psi _{n}(s),$ of \ the simple ansatz:

\begin{equation}
\psi _{n}(s)=\phi _{n}(s)y_{n}(s),
\end{equation}
reduces (1) into an equation of a hypergeometric type

\begin{equation}
\sigma (s)y_{n}^{\prime \prime }(s)+\tau (s)y_{n}^{\prime }(s)+\lambda
y_{n}(s)=0,
\end{equation}
where

\begin{equation}
\sigma (s)=\pi (s)\frac{\phi (s)}{\phi ^{\prime }(s)},
\end{equation}

\begin{equation}
\tau (s)=\widetilde{\tau }(s)+2\pi (s),\text{ }\tau ^{\prime }(s)<0,
\end{equation}
and $\lambda $ is a parameter defined as
\begin{equation}
\lambda =\lambda _{n}=-n\tau ^{\prime }(s)-\frac{n\left( n-1\right) }{2}%
\sigma ^{\prime \prime }(s),\text{ \ \ \ \ \ \ }n=0,1,2,....
\end{equation}
The polynomial $\tau (s)$ with the parameter $s$ and prime factors show the
differentials at first degree be negative. It is worthwhile to note that $%
\lambda $ or $\lambda _{n}$ are obtained from a particular solution of the
form $y(s)=y_{n}(s)$ which is a polynomial of degree $n.$ Further, the other
part $y_{n}(s)$ of the wave function (2) is the hypergeometric-type function
whose polynomial solutions are given by Rodrigues relation

\begin{equation}
y_{n}(s)=\frac{B_{n}}{\rho (s)}\frac{d^{n}}{ds^{n}}\left[ \sigma ^{n}(s)\rho
(s)\right] ,
\end{equation}
where $B_{n}$ is the normalization constant and the weight function $\rho
(s) $ must satisfy the condition [12]

\begin{equation}
\frac{d}{ds}w(s)=\frac{\tau (s)}{\sigma (s)}w(s),\text{ }w(s)=\sigma (s)\rho
(s).
\end{equation}
The function $\pi $ and the parameter $\lambda $ are defined as

\begin{equation}
\pi (s)=\frac{\sigma ^{\prime }(s)-\widetilde{\tau }(s)}{2}\pm \sqrt{\left(
\frac{\sigma ^{\prime }(s)-\widetilde{\tau }(s)}{2}\right) ^{2}-\widetilde{%
\sigma }(s)+k\sigma (s)},
\end{equation}
\begin{equation}
\lambda =k+\pi ^{\prime }(s).
\end{equation}
In principle, since $\pi (s)$ has to be a polynomial of degree at most one,
the expression under the square root sign in (9) can be arranged to be the
square of a polynomial of first degree [12]. This is possible only if its
discriminant is zero. In this case, an equation for $k$ is obtained. After
solving this equation, the obtained values of $k$ are substituted in (9). In
addition, by comparing equations (6) and (10), we obtain the energy
eigenvalues.

\section{Exact solutions of the quantum system with the NU method}

\label{ES}

\subsection{Seperating variables of the Schr\"{o}dinger equation}

The modified Kratzer potential plus ring-shaped potential in spherical
coordinates is defined as [47]

\begin{equation}
V(r,\theta )=D_{e}\left( \frac{r-r_{e}}{r}\right) ^{2}+\beta \frac{\cos
^{2}\theta }{r^{2}\sin ^{2}\theta },
\end{equation}
where $\beta $ is positive real constant. The potential in (11) introduced
by Cheng-Dai [47] reduces to the modified Kratzer potential in the limiting
case of $\beta =0$ [48]$.$ In fact the energy spectrum for this potential
can be obtained directly by considering it as special case of the general
non-central seperable potentials [45].

Our aim is to derive analytically the energy spectrum for a moving particle
in the presence of a potential (11) in a very simple way. The $D$%
-dimensional space Schr\"{o}dinger equation in spherical polar coordinates
written for potential (11) takes the form [1,6]

\[
-\frac{\hbar ^{2}}{2\mu }\left[ \frac{1}{r^{D-1}}\frac{\partial }{\partial r}%
\left( r^{D-1}\frac{\partial }{\partial r}\right) +\frac{1}{r^{2}}\left(
\frac{1}{\sin \theta }\frac{\partial }{\partial \theta }\left( \sin \theta
\frac{\partial }{\partial \theta }\right) +\frac{1}{\sin ^{2}\theta }\frac{%
\partial ^{2}}{\partial \varphi ^{2}}-\frac{2\mu \beta }{\hbar ^{2}}\frac{%
\cos ^{2}\theta }{\sin ^{2}\theta }\right) \right] \psi (r,\theta ,\varphi )
\]
\begin{equation}
+\left[ D_{e}\left( \frac{r-r_{e}}{r}\right) ^{2}-E\right] \psi (r,\theta
,\varphi )=0,
\end{equation}
where $\mu =\frac{m_{1}m_{2}}{m_{1}+m_{2}}$ being the reduced mass of the
two particles and $\psi (r,\theta ,\varphi )$ being the total wave function
separated as follows

\begin{equation}
\psi _{n\ell m}(r,\theta ,\varphi )=R(r)Y_{\ell }^{m}(\theta ,\varphi ),%
\text{ }R(r)=r^{-(D-1)/2}g(r),\text{ }Y_{\ell }^{m}(\theta ,\varphi
)=H(\theta )\Phi (\varphi ).
\end{equation}
On substituting equation (13) into (12) leads to a set of second-order
differential equations:
\begin{equation}
\left[ \frac{1}{r^{D-1}}\frac{d}{dr}\left( r^{D-1}\frac{d}{dr}\right) -\frac{%
L_{D-1}^{2}}{r^{2}}\right] R(r)+\frac{2\mu }{\hbar ^{2}}\left[ E-D_{e}\left(
\frac{r-r_{e}}{r}\right) ^{2}\right] R(r)=0,
\end{equation}

\begin{equation}
\left[ \frac{1}{\sin \theta }\frac{d}{d\theta }\left( \sin \theta \frac{d}{%
d\theta }\right) -\frac{m^{2}}{\sin ^{2}\theta }-\frac{2\mu \beta }{\hbar
^{2}}\frac{\cos ^{2}\theta }{\sin ^{2}\theta }+\ell (\ell +D-2)\right]
H(\theta )=0,
\end{equation}
\begin{equation}
\frac{d^{2}\Phi (\varphi )}{d\varphi ^{2}}+m^{2}\Phi (\varphi )=0,
\end{equation}
where $L_{D-1}^{2}=\ell (\ell +D-2).$ The solution in (16) is periodic and
must satisfy the period boundary condition $\Phi (\varphi +2\pi )=\Phi
(\varphi )$ from which we obtain
\begin{equation}
\Phi _{m}(\varphi )=\frac{1}{\sqrt{2\pi }}\exp (\pm im\varphi ),\text{ \ }%
m=0,1,2,.....
\end{equation}
Therefore, we are left to solve equations (14) and (15). After lengthy, but
straightforward, calculations, Equation (14), representing the radial wave
equation can be rewritten as [6]:
\begin{equation}
\frac{d^{2}g(r)}{dr^{2}}+\left[ \frac{2\mu }{\hbar ^{2}}(E-D_{e})+\frac{4\mu
D_{e}r_{e}}{\hbar ^{2}}\frac{1}{r}-\frac{\widetilde{\nu }+(2\mu
D_{e}r_{e}^{2}/\hbar ^{2})}{r^{2}}\right] g(r)=0,
\end{equation}
where
\begin{equation}
\widetilde{\nu }=\frac{1}{4}(M-1)(M-3),\text{ }M=D+2\ell .
\end{equation}
The two particles in equation (18) interacting via Coulombic-like field have
a slight change in the angular momentum barrier acts as a repulsive core
which for any arbitrary $\ell $ prevents collapse of the system in any space
dimension due to the additional centrifugal potential barrier. On the other
hand, equation (15) representing the angular wave equation takes the simple
form
\begin{equation}
\frac{d^{2}H(\theta )}{d\theta ^{2}}+\frac{\cos \theta }{\sin \theta }\frac{%
dH(\theta )}{d\theta }+\left[ \ell (\ell +D-2)-\frac{m^{2}+(2\mu \beta
/\hbar ^{2})\cos ^{2}\theta }{\sin ^{2}\theta }\right] H(\theta )=0.
\end{equation}
Therefore, equations (18) and (20) have to be solved latter by using the NU
method in the following subsections.

\subsection{The solutions of the angular equation}

In order to apply NU method, we introduce a new variable $s=\cos \theta ,$
equation (20) is then rearranged as the universal associated-Legendre
differential equation [45,49]

\begin{equation}
\frac{d^{2}H(s)}{ds^{2}}-\frac{2s}{1-s^{2}}\frac{dH(s)}{ds}+\frac{\nu
^{\prime }(1-s^{2})-m^{\prime }{}^{2}}{\sin ^{2}\theta }H(\theta )=0,
\end{equation}
where

\begin{equation}
\nu ^{\prime }=\ell ^{\prime }(\ell ^{\prime }+D-2)=\ell (\ell +D-2)+2\mu
\beta /\hbar ^{2}\text{ \ \ and \ \ \ }m^{\prime }{}^{2}=m^{2}+2\mu \beta
/\hbar ^{2}.
\end{equation}
The solution of this equation has already been solved by the NU method in
[45,47]. However, the aim in this subsection is to solve with different
parameters resulting from the $D$-space-dimensions of Schr\"{o}dinger
equation. Upon letting $D=3,$ we can readily obtain the standard case given
in [47]. Equation (21) is compared with (1) and the following
identifications are obtained

\begin{equation}
\widetilde{\tau }(s)=-2s,\text{ \ \ \ }\sigma (s)=1-s^{2},\text{ \ \ }%
\widetilde{\sigma }(s)=-\nu ^{\prime }s^{2}+\nu ^{\prime }-m^{\prime }{}^{2}.
\end{equation}
Inserting the above expressions into equation (9), one obtains the following
function:

\begin{equation}
\pi (s)=\pm \sqrt{(\nu ^{\prime }-k)s^{2}+k-\nu ^{\prime }+m^{\prime }{}^{2}}%
.
\end{equation}
Following the method, the polynomial $\pi (s)$ is found in the following
possible values
\begin{equation}
\pi (s)=\left\{
\begin{array}{cc}
m^{\prime }s & \text{\ for }k_{1}=\nu ^{\prime }-m^{\prime }{}^{2}, \\
-m^{\prime }s & \text{\ for }k_{1}=\nu ^{\prime }-m^{\prime }{}^{2}, \\
m^{\prime } & \text{\ for }k_{2}=\nu ^{\prime }, \\
-m^{\prime } & \text{\ for }k_{2}=\nu ^{\prime }.
\end{array}
\right.
\end{equation}
Imposing the condition $\tau ^{\prime }(s)<0,$ for equation (5), one selects

\begin{equation}
k_{1}=\nu ^{\prime }-m^{\prime }{}^{2}\text{ \ \ and \ \ }\pi (s)=-m^{\prime
}s,
\end{equation}
which yields form equation (5)
\begin{equation}
\tau (s)=-2(1+m^{\prime })s.
\end{equation}
Using equations (6) and (10), the following expressions for $\lambda $ are
obtained, respectively,

\begin{equation}
\lambda =\lambda _{n}=2n(1+m^{\prime })+n(n-1),
\end{equation}
\begin{equation}
\lambda =\nu ^{\prime }-m^{\prime }{}(1+m^{\prime }).
\end{equation}
We compare equations (28) and (29) and from the definition $\nu ^{\prime
}=\ell ^{\prime }(\ell ^{\prime }+D-2),$ the new angular momentum\ $\ell
^{\prime }$ values are obtained as

\begin{equation}
\ell ^{\prime }=-\frac{(D-2)}{2}+\frac{1}{2}\sqrt{(D-2)^{2}+4(n+\sqrt{%
m^{2}+2\mu \beta /\hbar ^{2}})(n+1+\sqrt{m^{2}+2\mu \beta /\hbar ^{2}})},
\end{equation}
which can be reduced to the simple form

\begin{equation}
\ell ^{\prime }=n+m^{\prime },
\end{equation}
in three-dimensions [47]. Using equations (2)-(4) and (7)-(8), the wave
function can be written as,

\begin{equation}
H_{m^{\prime }}(\theta )=N_{\ell ^{\prime }m^{\prime }}\sin (\theta
)^{m^{\prime }}P_{n}^{(m^{\prime },m^{\prime })}(\cos \theta ),
\end{equation}
where $N_{\ell ^{\prime }m^{\prime }}=\sqrt{\frac{(2\ell ^{\prime }+1)(\ell
^{\prime }-m^{\prime })!}{2(\ell ^{\prime }+m^{\prime })!}}$ is the
normalization constant given in [46,47] and

\begin{equation}
n=-\frac{(1+2m^{\prime })}{2}+\frac{1}{2}\sqrt{(2\ell ^{\prime
}+1)^{2}+4\ell ^{\prime }(D-3)},
\end{equation}
with $m^{\prime }$ is defined by equation (22).

\subsection{The solutions of the radial equation}

The solution of the SE for the modified central Kratzer's potential has
already been solved by means of NU-method in [48]. Very recently, using the
same method, the problem for the non-central potential in (11) has been
solved in three dimensions by Cheng and Dai [47]. However, the aim of this
subsection is to solve the problem with a different radial separation
function $g(r)$ in any arbitrary dimensions. We now study the bound-states
(real) solution $E<D_{e}$ of equation (18). Letting
\begin{equation}
\varepsilon =\sqrt{-\frac{2\mu }{\hbar ^{2}}(E-D_{e})},\text{ }\alpha =\frac{%
4\mu D_{e}r_{e}}{\hbar ^{2}},\text{ }\gamma =\widetilde{\nu }+\frac{1}{2}%
\alpha r_{e},
\end{equation}
and substituting these expressions in equation (18), one gets
\begin{equation}
\frac{d^{2}g(r)}{dr^{2}}+\left( \frac{-\varepsilon ^{2}r^{2}+\alpha r-\gamma
}{r^{2}}\right) g(r)=0.
\end{equation}
To apply the conventional NU-method, equation (35) is compared with (1) and
the following expressions are obtained

\begin{equation}
\widetilde{\tau }(r)=0,\text{ \ \ \ }\sigma (r)=r,\text{ \ \ }\widetilde{%
\sigma }(r)=-\varepsilon ^{2}r^{2}+\alpha r-\gamma .
\end{equation}
Substituting the above expressions into equation (9) gives

\begin{equation}
\pi (r)=\frac{1}{2}\pm \frac{1}{2}\sqrt{4\varepsilon ^{2}r^{2}+4(k-\alpha
)r+4\gamma +1}.
\end{equation}
According to this conventional method, the expression in the square root be
the square of a polynomial. Thus, the two roots $k$ can be readily obtained
as

\begin{equation}
k=\alpha \pm \varepsilon \sqrt{4\gamma +1}.
\end{equation}
In view of that, we arrive at the following four possible functions of $\pi
(r):$%
\begin{equation}
\pi (r)=\left\{
\begin{array}{cc}
\frac{1}{2}+\left[ \varepsilon r+\frac{1}{2}\sqrt{4\gamma +1}\right] & \text{%
\ for }k_{1}=\alpha +\varepsilon \sqrt{4\gamma +1}, \\
\frac{1}{2}-\left[ \varepsilon r+\frac{1}{2}\sqrt{4\gamma +1}\right] & \text{%
\ for }k_{1}=\alpha +\varepsilon \sqrt{4\gamma +1}, \\
\frac{1}{2}+\left[ \varepsilon r-\frac{1}{2}\sqrt{4\gamma +1}\right] & \text{%
\ for }k_{2}=\alpha -\varepsilon \sqrt{4\gamma +1}, \\
\frac{1}{2}-\left[ \varepsilon r-\frac{1}{2}\sqrt{4\gamma +1}\right] & \text{%
\ for }k_{2}=\alpha -\varepsilon \sqrt{4\gamma +1}.
\end{array}
\right.
\end{equation}
The correct value of $\pi (r)$ is chosen such that the function $\tau (r)$
given by equation (5) will have negative derivative [12]. So we can select
the physical values to be

\begin{equation}
k=\alpha -\varepsilon \sqrt{4\gamma +1}\text{ \ \ and \ \ }\pi (r)=\frac{1}{2%
}-\left[ \varepsilon r-\frac{1}{2}\sqrt{4\gamma +1}\right] ,
\end{equation}
which yield
\begin{equation}
\tau (r)=-2\varepsilon r+(1+\sqrt{4\gamma +1}).
\end{equation}
Using equations (6) and (10), the following expressions for $\lambda $ are
obtained, respectively,

\begin{equation}
\lambda =\lambda _{n}=2N\varepsilon ,\text{ }N=0,1,2,...,
\end{equation}
\begin{equation}
\lambda =\alpha -\varepsilon (1+\sqrt{4\gamma +1}).
\end{equation}
So we can obtain the energy eigenvalues as

\begin{equation}
E_{N}=D_{e}-\frac{8\mu D_{e}^{2}r_{e}^{2}/\hbar ^{2}}{\left( 2N+1+\sqrt{%
(M-1)(M-3)+8\mu D_{e}r_{e}^{2}/\hbar ^{2}+1}\right) ^{2}},\text{ }
\end{equation}
where

\begin{equation}
(M-1)(M-3)=4\widetilde{\nu }=(D-2)^{2}+4\ell ^{\prime }(\ell ^{\prime
}+D-2)-8\mu \beta /\hbar ^{2}-1,
\end{equation}
with $\ell ^{\prime }$ defined in (30). Therefore, the final energy spectra
in equation (44) take the following Coulombic-like form [7]
\begin{equation}
E_{N^{\prime }}=D_{e}-\frac{2\mu D_{e}^{2}r_{e}^{2}/\hbar ^{2}}{\left(
N^{\prime }\right) ^{2}},\text{ }N^{\prime }=0,1,2,...
\end{equation}
where

\begin{equation}
N^{\prime }=\frac{1}{2}\left[ 2N+\sqrt{(D-2)^{2}+4\ell ^{\prime }(\ell
^{\prime }+D-2)+8\mu (D_{e}r_{e}^{2}-\beta )/\hbar ^{2}}+1\right] ,
\end{equation}
is simply obtained by means of substituting equation (45) into (44).

(i) If $D=3$, equation (44), with the help of equation (45), is transformed
into the following form

\begin{equation}
E_{Nnm}=D_{e}-\frac{8\mu D_{e}^{2}r_{e}^{2}/\hbar ^{2}}{\left( 2N+1+\sqrt{%
(2n+1)^{2}+4m^{2}+4(2n+1)\sqrt{m^{2}+2\mu \beta /\hbar ^{2}}+8\mu
D_{e}r_{e}^{2}/\hbar ^{2}}\right) ^{2}},
\end{equation}
and it is consistent with equation (40) in [47].

(ii) If, $D=3$ and $\beta =0$ (modified Kratzer potential)$,$ equation (44)
is transformed into the form

\begin{equation}
E_{n}=D_{e}-\frac{8\mu D_{e}^{2}r_{e}^{2}/\hbar ^{2}}{\left( 1+2n+\sqrt{%
1+4\ell (\ell +1)+8\mu D_{e}r_{e}^{2}/\hbar ^{2}}\right) ^{2}}.\text{ }
\end{equation}
and it is consistent with equation (14) in [48].

Let us now turn attention to find the radial wavefunctions for this
potential. Using $\tau (r),$ $\pi (r)$ and $\sigma (r)$ in equations (4) and
(8), we find
\begin{equation}
\phi (r)=r^{(\sqrt{4\gamma +1}+1)/2}e^{-\varepsilon r},
\end{equation}

\begin{equation}
\rho (r)=r^{\sqrt{4\gamma +1}}e^{-2\varepsilon r}.
\end{equation}
Then from equation (7), we obtain

\begin{equation}
y_{n}(r)=B_{n}r^{-\sqrt{4\gamma +1}}e^{2\varepsilon r}\frac{d^{N}}{dr^{N}}%
\left( r^{N+\sqrt{4\gamma +1}}e^{-2\varepsilon r}\right) ,
\end{equation}
and the wave function $g(r)$ can be written in the form of the generalized
Laguerre polynomials as

\begin{equation}
g(r)=C_{N,L}r^{L+1}e^{-\varepsilon r}L_{N}^{2L+1}(2\varepsilon r),
\end{equation}
where $L$ can be found easily from equation (47). Finally, the radial wave
functions of the Schr\"{o}dinger equation are obtained
\begin{equation}
R(r)=C_{N,L}r^{L-(D-3)/2}e^{-\varepsilon r}L_{N}^{2L+1}(2\varepsilon r),
\end{equation}
where

\begin{equation}
\varepsilon =\frac{\mu a}{\hbar ^{2}N^{\prime }}
\end{equation}
with $N^{\prime }$ is given in equation (47) and $C_{N,L}$ is the
normalization constant to be determined below. Using the normalization
condition, $\int\limits_{0}^{\infty }R^{2}(r)r^{D-1}dr=1,$ and the
orthogonality relation of the generalized Laguerre polynomials, $%
\int\limits_{0}^{\infty }z^{\eta +1}e^{-z}\left[ L_{n}^{\eta }(z)\right]
^{2}dz=\frac{(2n+\eta +1)(n+\eta )!}{n!},$ we have

\begin{equation}
C_{N,L}=\sqrt{\frac{(2\varepsilon )^{2L+3}N!}{2(N+L+1)(N+2L+1)!}}.
\end{equation}
Therefore, we may express the normalized total wave functions as

\[
\psi (r,\theta ,\varphi )=\sqrt{\frac{(2\varepsilon )^{2L+3}(2\ell ^{\prime
}+1)(\ell ^{\prime }-m^{\prime })!N!}{4\pi (\ell ^{\prime }+m^{\prime
})!(N+L+1)(N+2L+1)!}}r^{L-(D-3)/2}\exp (-\varepsilon
r)L_{N}^{2L+1}(2\varepsilon r)
\]
\begin{equation}
\times \sin (\theta )^{m^{\prime }}P_{n}^{(m^{\prime },m^{\prime })}(\cos
\theta )\exp (\pm im\varphi ).
\end{equation}

\section{Results and Conclusions}

\label{RAC}In this paper, the Schr\"{o}dinger equation in any
arbitrary dimensions has been solved for its exact bound-states with
a recently proposed modified Kratzer potential plus ring-shaped
potential by means of a simple conventional NU method. The
analytical expressions for the total energy levels of this system is
found to be different from the results obtained for the modified
Kratzer's potential in [48] and also more general than the one
obtained recently in three-dimensions [47]. Therefore, the
noncentral potentials treated in [45] can be introduced as
perturbation to the modified Kratzer's potential by adjusting the
strength of the coupling constant $\beta $ in terms of $D_{e},$
which is the coupling constant of the modified Kratzer's potential.
In addition, the angular part, the radial part and then the total
wave functions are also found. Thus, the Schr\"{o}dinger equation
with a new non-central but separable potential has also been studied
(cf. [45] and the references therein). This method is very simple
and useful in solving other complicated systems analytically without
given a restiction conditions on the solution of some quantum
systems as the case in the other models. Finally, we point out that
these exact results obtained for this new proposed form of the
potential (11) may have some interesting applications in the study
of different quantum mechanical systems, atomic and molecular
physics.

\acknowledgments This research was partially supported by the
Scientific and Technological Research Council of Turkey. S.M.
Ikhdair wishes to dedicate this work to his family for their love
and assistance.

\end{document}